\documentclass[11pt]{article} 
\pdfoutput=1

\usepackage{color}
\usepackage{pslatex}
\usepackage{pdfpages}
\usepackage{epsfig}
\usepackage{graphicx}
\usepackage{psfrag}
\usepackage{pdftricks}
\usepackage{multirow}
\usepackage{graphicx}
\usepackage{hyperref}

\def\baselinestretch{1.3}
\newcommand{\ba}{\begin{array}}
\newcommand{\ea}{\end{array}}
\newcommand{\bd}{\begin{displaymath}}
\newcommand{\ed}{\end{displaymath}}
\newcommand{\be}{\begin{equation}}
\newcommand{\ee}{\end{equation}}
\newcommand{\bea}{\begin{eqnarray}}
\newcommand{\eea}{\end{eqnarray}}
\newcommand{\sla}[1]{/\!\!\!#1}

\def\a{\alpha}

\def\q2 {q^2}

\def\r {\rightarrow}

\def\miss {\hspace{-0.4cm}\slash~~}

\def\neu {\tilde{\chi}^0}
\def\rslep {\tilde{e_R}}

\def\snu {\tilde{\nu}}
\def\lslep {\tilde{e_L}}
\def\stau {\tilde{\tau}}
\def\mer {m_{\rslep}}
\def\mmr {m_{\tilde{\mu}_R}}
\def\mml {m_{\tilde{\mu}_L}}
\def\mel {m_{\lslep}}

\def\bt{\begin{table}}
\def\et{\end{table}}

\catcode`@=11 
\def \gsim{\mathrel{\mathpalette\@versim>}}
\def \lsim{\mathrel{\mathpalette\@versim<}}
\def \@versim#1#2{\lower0.4ex\vbox{\baselineskip\z@skip\lineskip\z@skip
     \lineskiplimit\z@\ialign{$\m@th#1\hfil##\hfil$%
     \crcr#2\crcr\sim\crcr}}}
\catcode`@=12 

\begin{document}

\title{Light neutralino dark matter in the MSSM and its implication for  LHC searches for staus}

\author{Genevi\`eve B\'elanger$^1$, Sanjoy Biswas$^{2}$, C\'eline B\oe hm$^{1,3}$, Biswarup Mukhopadhyaya$^4$}

\maketitle

\noindent 
{$^1$ LAPTH, Univ. de Savoie, CNRS, B.P.110, F-74941 Annecy-le-Vieux Cedex, France \\ 
$^2$ INFN, Sezione di Roma, Dipartimento di Fisica, Universit\`a di Roma La Sapienza,
Piazzale Aldo Moro 2, I-00185 Rome, Italy. \\
$^3$ Institute for Particle Physics Phenomenology, University of Durham, Durham, DH1 3LE, UK. \\
$^4$   Regional Centre for Accelerator-based Particle Physics,
            Harish-Chandra Research Institute, 
            Chhatnag Road, Jhusi, Allahabad - 211019, India.
}

\begin{abstract} 
It was shown in a previous study that a lightest neutralino with mass below 30 GeV was severely constrained in the minimal supersymmetric standard model (MSSM), unless it annihilates via a light stau and thus yields the observed  dark matter abundance. In such a scenario, while the stau is the next-to-lightest supersymmetric particle (NLSP), the charginos and the other neutralinos as well as sleptons of the first two families are also likely to be not too far above the mass bounds laid down by the Large Electron Positron (LEP) collider. As the branching ratios of decays of the charginos and the next-to-lightest neutralino into staus are rather large, one expects
significant rates of tau-rich final states in such a case. With this in view, we investigate the same-sign ditau and tri-tau signals of this scenario at the Large Hadron Collider (LHC) for two MSSM benchmark points corresponding to light neutralino dark matter. The associated signal rates for these channels 
are computed, for the centre-of-mass energy of 14 TeV. We find that both channels lead to appreciable rates if the squarks and the gluino
are not too far above a TeV, thus allowing to probe scenarios with light neutralinos  in the 14 TeV LHC run with $10-100 {\rm  fb}^{-1}$.
\end{abstract}


{\small \footnoterule{$^{\dagger}$ e-mail: belanger@lapp.in2p3.fr, sanjoy.biswas@roma1.infn.it, c.m.boehm@durham.ac.uk, biswarup@hri.res.in}}

\newpage
\setcounter{footnote}{0}

\def\baselinestretch{1.5}
\section{Introduction}

 Searches for scenarios beyond the standard model, which offer
solutions to the 'naturalness issue and/or the dark matter problem,
has been one of the main goals of the large hadron collider (LHC).  At
the end of its operation at 7 TeV, leading to accumulated data
corresponding to about 5 $fb^{-1}$ of integrated luminosity, the LHC
has seen hints (though contentious) of a Higgs particle with a mass
near 125 GeV; however, no signal of new particles, including a
potential dark matter candidate, has yet revealed itself.  In view of
this, a large part of the parameter space of the most studied
extension of the standard model, namely, the constrained minimal
supersymmetric (SUSY) standard model (CMSSM), has been excluded.  In
particular, lower limits for squarks and gluinos in such a scenario
has crossed the 1 TeV mark~\cite{Aad:2011ib,Aad:Moriond,CMS_squark,Chatrchyan:2011zy}. Within the framework of the CMSSM, these limits in turn  imply constraints on  sleptons and electroweak gauginos whose masses are correlated with those of the strong sector.  On the other hand, in the more general minimal SUSY standard model (MSSM) with free parameters defined at the electroweak scale, there are no direct correlations between the parameters of the strong and electroweak sector. Furthermore the constraints on the coloured sector  from the LHC data can be somewhat relaxed. Thus the possibility of  relatively light neutralinos, charginos or sleptons  cannot yet be ruled out on the basis of existing data. In particular, the lightest neutralino as the lightest supersymmetric particle (LSP), at or below the electroweak scale, continues to offer itself as the dark matter (DM) candidate.

Early interest in supersymmetric models with light neutralinos~\cite{Bottino:2002ry,Hooper:2002nq,Belanger:2003wb,Boehm:2002yz} was renewed  by the hints of possible signals in direct detection experiments, 
DAMA~\cite{Bernabei:2008yi,Bernabei:2010mq}, CoGeNT~\cite{Aalseth:2010vx}, Cresst~\cite{Angloher:2011uu}. However,
new particles below the electroweak scale are constrained 
 both by collider searches for the Higgs and new particles (LEP, Tevatron and LHC), B-physics observables, precision measurements, as well as astrophysics constraints. 
Nevertheless, several studies have shown that within the MSSM with non-universal gaugino masses, a neutralino LSP with a mass 
smaller than 30 GeV (down to ${\cal O}(10) \ {\rm GeV}$) could satisfy 
all constraints~\cite{Vasquez:2010ru, Belikov:2010yi, Dreiner:2009ic, Cumberbatch:2011jp,Fornengo:2010mk,Calibbi:2011ug}. After applying LEP limits on charged particles, the most stringent constraint
 on the light neutralino (dominantly Bino) comes from the upper bound on the dark matter relic density measured by WMAP which requires an efficient mechanism for Bino annihilation.
One such mechanism corresponds to Bino annihilation into fermion pairs via t-channel exchange of a slepton, with the largest  annihilation cross section when the slepton is light. 
Universality of the soft masses for sleptons and a large mixing in the third generation, with a large value of $\tan\beta$, the
ratio of the two Higgs vacuum expectation values (vev), implies that the lightest slepton is the stau which therefore gives the largest contribution 
to neutralino annihilation. Furthermore the lower bound from LEP is only  $m_{\tilde\tau}>81.9$ GeV.
Other constraints from direct detection of dark matter and from observations of gamma-rays from dark matter annihilation in the galaxy play an important role 
~\cite{Vasquez:2010ru,AlbornozVasquez:2011yq} as well as limits on the Higgs sector and on B-observables from the LHC~\cite{bsmu_CMSLHCB}.

A light stau therefore buttresses the light neutralino scenario from
the dark mater point of view, and it is obviously interesting to look
for hitherto unexplored signatures of such a scenario.
The largest rate for production of staus comes from the decay chains of the strongly interacting 
squarks and gluinos. 
In particular, the scenario with a light Bino LSP which we will analyse has in addition light Higgsinos,  as this 
favours large  branching ratios of the chargino and neutralino decaying into staus. 
We therefore expect final states with many tau leptons since the lightest stau mainly decays into a $\tilde{\chi}_1^0 \tau$ pair. 

 In proposed leptonic signals of SUSY, same-sign dileptons as well as
trileptons have often been advocated as relatively clean. However,
similar signals with taus have not received that much attention as
yet, though the identification of tau-charge in its one-prong decay is
a distinct possibility now. Also, in cases where a high rate of tau
production is archetypal of a specific scenario, it makes sense to look for tri-tau events, where detectable rates may be salvaged
despite suppressions due to branching fraction and tau detection efficiency. The specific signature we
will consider include same-sign di-tau or tri-tau final states
associated with hard jets and missing transverse energy ($\sla{E_T}$).
Some earlier studies mainly in the context of gauge mediated SUSY breaking involving tau leptons
can be found in \cite{Dicus:1997yp,Dutta:1998tt,Muller:1998hw,Baer:2000pe}. An early 
study to limit the minimal supergravity parameter space using the Tevatron data can be found in \cite{Abazov:2009rj}.

 Since the final states of our concern depends crucially on cascades
triggered by squark and gluino production, the event rates will depend
critically on the mass of the squarks/gluinos.  Here we have chosen
benchmark points where these masses are well above a TeV. Thus, not
only they are consistent with existing experimental constraints, but
our predictions are also on the conservative side.
 
The paper is organised as follows. In Section 2 we recall the conditions for having light neutralinos in the MSSM and identify benchmark points representative of 
light neutralino scenarios and which are expected to lead to a large tau production. The set-up of the analysis  is described in Section 3 and the results are presented 
in Section 4. We summarise and conclude in Section 5.

\section{Scenarios with large $\tau$ production}

In this section, we focus on the production and decay channels which lead to a large $\tau$ production and identify two benchmark scenarios in order to perform 
a quantitative study of the di and tri-$\tau$ signatures expected at LHC in the case of neutralinos lighter than $\sim$ 30 GeV. We begin
  the discussion on this by emphasizing the viability of a light neutralino 
  and a stau just above it, from the viewpoint of the dark matter content of 
  the universe.

\subsection{A light neutralino and its annihilation}
We start by summarising the results obtained in Ref.~\cite{AlbornozVasquez:2011yq}. In this reference a Markov Chain Monte Carlo (MCMC) was used to explore the MSSM parameter 
space in order to pinpoint allowed masses
  of the lightest possible neutralino in the $[\sim 1,\sim 30]$ GeV range in light of the dark matter relic density requirement as well as astrophysical, 
flavour and collider constraints. On the whole,
  eleven free parameters are considered, all defined at the electroweak scale, namely
\begin{equation}
M_1,M_2,M_3,\mu, \tan\beta, M_A,  M_{\tilde l_R},M_{\tilde l_L}, M_{\tilde Q_{1,2}}, M_{\tilde Q_3}, A_t.
\end{equation}
Although the trilinear couplings
  $A_b$ and $A_\tau$ are both set to zero, the choice of large $\tan\beta$
  enables one to have appreciable left-right mixing in the  sbottom and
  stau sectors, due to the non negligible $b$ and 
$\tau$ mass and the fact that such mixing is proportional to $A_\tau-\mu\tan\beta$. Since the relic density constraint requires efficient neutralino annihilations, 
two dominant mechanisms emerge for getting an abundance of light (sub 30 GeV) neutralinos compatible with the observed dark matter cosmological parameter: (A) annihilations into 
lepton pairs through slepton exchange, (B) annihilation via light pseudoscalar Higgs exchange.

Scenario (A) relies on a Bino LSP and light sleptons. 
The lighter the exchange slepton is, the larger is the annihilation cross-section. This can be achieved by considering large 
values of $\tan \beta$ since this induces a large mixing in the stau sector in particular and thus decreases the mass of the lightest stau 
(i.e. increases the neutralino annihilation cross section into tau pairs). 
As a result, stau masses just above the LEP bound and up to 200 GeV were favoured by the MCMC to provide the correct neutralino relic density. 
Note that very large mixing in the stau sector, as induced by a large value for $\mu$,  can lead to the lightest stau below the LEP bound, thus 
the parameter $\mu$ has to be moderately small ($\simeq$ 250 GeV). 
This  implies relatively low masses for $\tilde\chi^{\pm}_1$ and $\tilde\chi^0_{2,3}$. Such 
 a situation yields a mass of the lighter stau in the desired range, and 
 at the same time lends the requisite Higgsino component to the lightest
 neutralino state.

In fact, the right-handed smuons and selectrons,
 too, contribute significantly to the annihilation rate of the LSP, even
 though they are not as light as the staus due to the 
lack of mixing (and despite a common soft mass for the three 
generations of sleptons). The reason is that right-handed components couple more strongly to the Bino. 
Thus a typical spectrum for scenario A consists of several light particles in addition to the LSP, namely the sleptons 
$\tilde\tau_1,\tilde{e}_R,\tilde\mu_R$ and gauginos $\neu_1, \neu_2, \neu_3, \tilde\chi^{\pm}_1$. The rest of our analysis will actually be based 
on this property.

Unlike scenario (A), the scenario (B) requires a neutralino LSP with as large a Higgsino component as  possible so as to ensure large couplings to the the neutral pseudoscalar $A^0$. 
This property holds even if the LSP is dominantly a Bino, implying in this case small values of the $\mu$ parameter since $M_1\ll \mu$. This scenario requires in addition 
a light pseudoscalar Higgs boson  as well as large values of tan$\beta$ to enhance the couplings of the Higgses to fermions.  However the low $M_A$ - large $\tan\beta$ region 
is strongly constrained by Tevatron and LHC searches~\cite{Chatrchyan:2011nx}. 

In what follows, we will actually disregard scenario (B), since 
 it was found in reference \cite{AlbornozVasquez:2011yq} that not only it is subject to the aforementioned constraints from the LHC, but also that the predicted direct detection rates and the photon flux from dark matter annihilation in Dwarf Spheroidal Galaxies were in excess of the limits from Xenon100 and FermiLAT respectively. Thus we focus exclusively on scenario (A) and investigate its signatures in the di and tri-tau final states, when one stau mass eigenstate is light enough.

\subsection{Tau production channels}

The largest production cross sections at the LHC are obtained for squarks and gluinos. These coloured particles decay into quarks and charginos/neutralinos which 
further decay directly or through a multistep decay chain  into the LSP and Standard Model fermions. 

As mentioned in the previous section, the spectrum associated with the scenario A involves a light stau NLSP (here by light we 
mean ${\cal O}(100) \ {\rm GeV}$ as well as light $\tilde\chi_{2,3}^0, \tilde\chi_1^{\pm}$ and right-handed slectrons and smuons.  
In the configurations where both the $\tilde\chi_{2,3}^0, \tilde\chi_1^{\pm}$  are heavier than the $\tilde{e}_R$ and $\tilde{\mu}_R$, 
the next-to-lightest neutralinos can directly decay into these light right-handed sfermion states, however, since these decays occur
via the bino component of $\tilde\chi^0_2$ or through the left-handed admixture in the lighter slepton mass eigenstate (which is almost
negligible for first two generations) the corresponding decay branching ratios are also very small. Hence, the next-to-lightest neutralino dominantly
decays into the $\tilde{\tau} \tau$, $\tilde\chi_1^{\pm} W^*$ and $\tilde\chi_{1,2}^0 Z^{(*)}$ channels, thus producing one or two taus in the final 
state. The lightest chargino instead predominantly decays into $\tilde\chi_{1,2}^0 W^{(*)}$, $ \tilde\tau \nu_\tau$ and $ \tilde\nu_\tau \tau$, 
thus producing one tau in the final state. Note that the $\tilde\chi_1^{\pm}$ decay rates into pure right-handed selectrons and smuons are also suppressed 
due to the large Higgsino fraction associated with the chargino.

The production of  
$\tilde\chi_i^0 $ and $\tilde\chi_1^{\pm}$  from each  squark and gluinos decay chains in $\tilde q \tilde q^{(*)},\tilde g\tilde g,\tilde q \tilde g$ processes can therefore give rise to di-tau as well as tri-tau signatures.  Other modes of tau production are in fact possible, for example from electroweak production of charginos, neutralinos and staus,  but in the following we will show that after cuts the squark/gluino initiated processes are the main channels to consider
\footnote{
Di-tau signatures of light staus from electroweak production of $\tilde\tau_1\tilde\tau_1$ and $\tilde\tau_1 \tilde\nu_\tau$ were  investigated in ~\cite{Carena:2012gp} for the LHC 14 TeV with a  luminosity of $200{\rm fb}^{-1}$. }. The question that we want to address in this paper is whether one can probe the existence of light MSSM neutralinos by using the di-taus (same-sign or opposite sign) or  tri-taus signatures at the LHC.

\begin{figure}[ht]
\includegraphics[width=15cm]{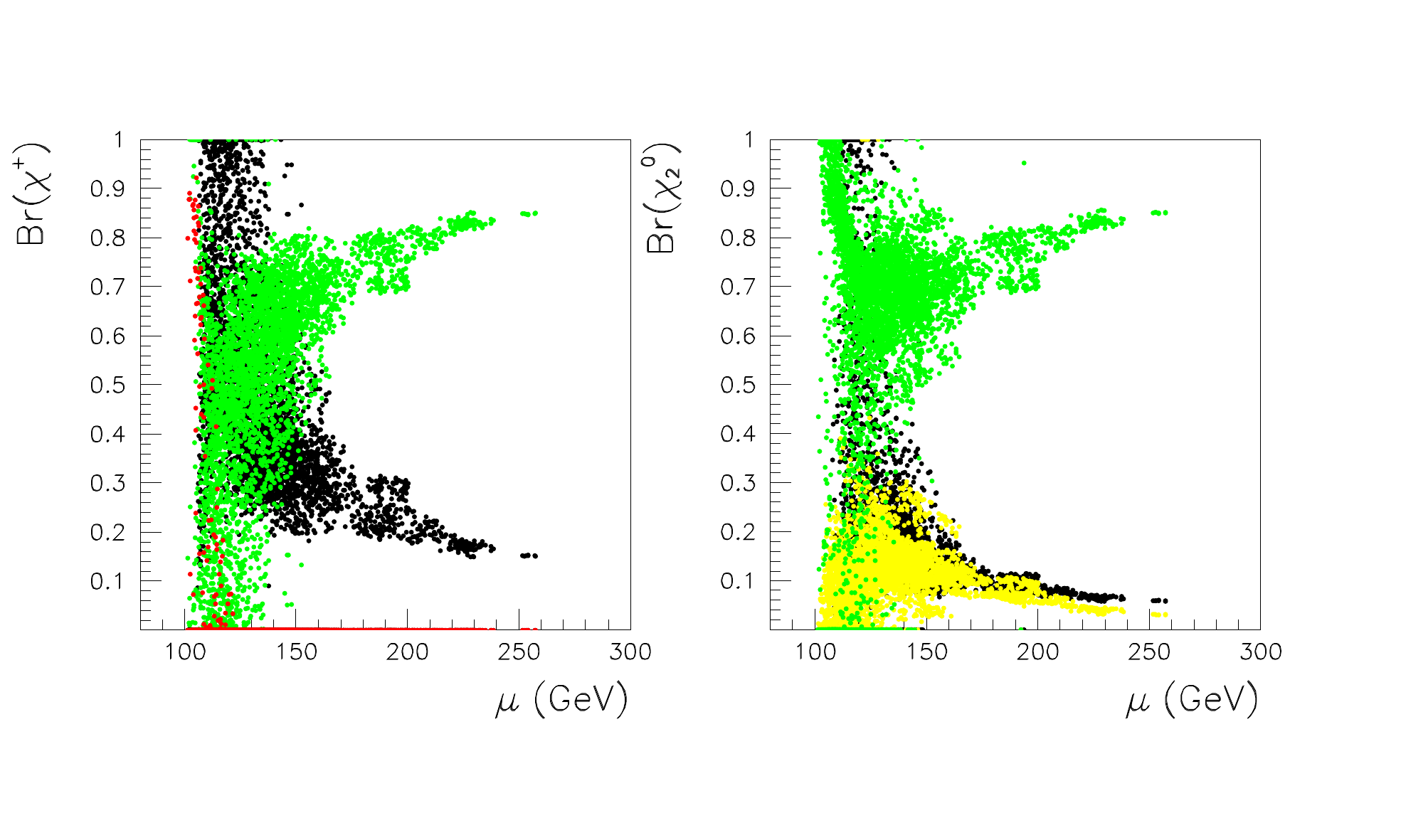}
\vskip 10pt
\vspace*{-1cm}
\caption{\small \it { Branching ratios of $\tilde\chi^{+}_1$ and $\tilde\chi^{+}_2$ as a function of $\mu$ for the MCMC points of ~\cite{AlbornozVasquez:2011yq}, left panel: $Br(\tilde\chi^+\rightarrow \tilde\tau\nu_\tau)$(green) $Br(\tilde\chi^+\rightarrow W\tilde\chi_1)$ (black) $Br(\tilde\chi^+\rightarrow\tilde\nu\tau)$(red) , 
 right panel:   $Br(\tilde\chi^0_2\rightarrow \tilde\tau\tau)$ (green)$Br(\tilde\chi^0_2\rightarrow Z\tilde\chi_1)$ (black),
 $Br(\tilde\chi^0_2\rightarrow \tilde\nu l)$(yellow).
}} 
\label{fig:channels}
\end{figure}

To answer this question,  we first need an estimate of the yield into taus that is expected in the models corresponding to the case (A); hence  the need to compute the chargino and neutralino decay modes for the majority of the points selected by the MCMC. We  found that 
for a large fraction of the points, the dominant branching ratios correspond to  Br$(\tilde\chi^{\pm}_1\r\stau_1\nu_{\tau}$) and Br$(\neu_2\r\stau\tau)$, as illustrated in Fig.~\ref{fig:channels}, meaning that taus are mostly produced through staus decay. 
 The main competing decay channels for charginos and neutralinos are into gauge bosons, these branching fraction are largest for values of $\mu<150 \ {\rm GeV}$   as shown in Fig.~\ref{fig:channels}. The gauge boson decay will contribute to the production of taus with a   small branching fraction. 
The $\tilde\chi^{\pm}_1\r\snu_{\tau}\tau$ decay can be large sometimes and therefore also contribute to getting taus in the final state.  Finally we note that the branching ratios of the second next-to-lightest neutralinos into the first and second generations of sleptons can reach up to 30\%, due to the sub-dominant Bino component of $\tilde\chi_2$, thus reducing the yield of tau channels that can be expected from these states. 

\begin{table}[htbp]
\begin{tabular}{||c|c|c||c||c|c||}
\hline
\hline
\multicolumn{3}{||c||}{Inputs parameters} & Particles  & {\bf BP-1}&{\bf BP-2}\\
\hline
 $M_1$ &  32.8  & 35.6   &  $\mel,\mml$  & 565  & 181.3  \\
 $M_2$ &  166  &  1487  &  $\mer,\mmr$  & 104 & 101.9   \\
 $M_3$ &  891.1  & 1400   &  $m_{\snu_{e_L}},m_{\snu_{\mu_L}}$& 560 & 163.5 \\
 $\mu$ &  145  &  110.7  &  $m_{\snu_{\tau_L}}$ & 560 & 163.5  \\
 $\tan\beta$ & 50.7  &  31.9  &     $m_{\stau_1}$& 101 & 93.8  \\
 $M_A$       &  690  & 1190   &   $m_{\stau_2}$& 566 & 185.6 \\
 $M_{\tilde{l}_L}$ &  564   &  175.5   &    $m_{\chi^0_1}$ & 28.8 & 28.8 \\
 $M_{\tilde{l}_R}$ &  94.2  &  91.7    &  $m_{\chi^0_2}$& 112.9 & 121.6  \\
 $M_{\tilde{q}_L}$ &  1133  &  1815    &  $m_{\chi^0_3}$& 161.8 & 123.5  \\
 $M_{\tilde{q}_R}$ &  1133  &  1815    &  $m_{\chi^0_4}$& 226.8 & 1500.9  \\
 $M_{\tilde{q}_{3L}}$ & 1318   &  1662    &  $m_{\chi^{\pm}_1}$& 111.8 & 113  \\
 $M_{\tilde{t}_R}$ &  1318  &  1300    &  $m_{\chi^{\pm}_2}$& 228 & 1500.9  \\
 $M_{\tilde{b}_R}$ &  1318  &  1100    &   $m_{\tilde{g}}$& 1007.8 & 1529.8 \\
 $A_t$             & -942   & -400     &     $m_{\tilde{t}_1}$& 1280.8 & 1343.6 \\
 $A_b$             &  0     &   0       &  $m_{\tilde{t}_2}$& 1391.5 & 1714.4  \\
 $A_{\tau}$        &   0    &   0       &  $m_{\tilde{d}_L}$& 1168.7 & 1854  \\
                   &   &   &      $m_{\tilde{d}_R}$& 1167.5 & 1853  \\%

\cline{1-3}
 \multicolumn{1}{||r}{Branching} & \multicolumn{1}{l}{fractions} &  \multicolumn{1}{c||}{}  
&  $m_{\tilde{u}_L}$& 1166 &  1852 \\
\cline{1-3}

 Br$(\neu_2\r\stau_1\tau)$  &  0.84 &  0.81  &    $m_{\tilde{u}_R}$& 1166.7 & 1852.7  \\
 Br$(\tilde{\chi}^+_1\r\stau^+_1\nu_{\tau})$  &  0.63  & 0.53  &  $m_{h^0}$ & 117 & 114.9 \\
 Br$(\neu_3\r\stau_1\tau)$  &  0.71 &  0.56  &    & &  \\
\hline
\hline
\end{tabular}\\

\begin{minipage}{5 in}
\caption {\small \it Proposed benchmark points (BP) for the study of MSSM with 
light neutralino dark matter including the branching fractions of lightest chargino
and second and third lightest neutralino. All the parameters having dimension of mass are given 
in GeV.}
\label{tab:1}      
\end{minipage}
\end{table}

\subsection{Benchmark points}

In order to perform a more quantitative study, we shall now identify  two benchmark points which are representative of the features and spectrum of scenario A. For selection, we require that both charginos and neutralinos have a large branching fraction into taus. Their characteristics are summarised in table \ref{tab:1}. Note that their mass spectrum is obtained using the spectrum 
generator {\bf \begin{footnotesize} SuSpect 2.41\end{footnotesize}} \cite{Djouadi:2002ze}.

Not only are these benchmark points compatible with the WMAP data \cite{Komatsu:2008hk} and other DM search experiments, but they are also 
consistent with experimental constraints such as those from $b\rightarrow s\gamma$, $B_s\rightarrow \mu^+\mu^-$, correction to the $\rho$-parameter and muon 
($g - 2$) \cite{Amsler:2008zzb,Nakamura:2010zzi,Barberio:2008fa,AlbornozVasquez:2011yq}. Furthermore we have checked using {\tt HiggsBounds3.6.1}~\cite{Bechtle:2008jh,Bechtle:2011sb} that both benchmark points are compatible with recent LHC limits on the Higgs~\cite{ATLAS:2012ad,ATLAS:2012ae,Chatrchyan:2012tw,Chatrchyan:2012tx}. 
The first benchmark point, hereafter referred to as BP-1 lies within the mass window still allowed for a SM Higgs, while the second point --hereafter referred to as BP-2-- has a suppressed signal strength as compared to the SM. 

Note that the exact mass of the light Higgs is not critical for this analysis. Equivalent benchmark points with a Higgs mass near $125$ GeV can be obtained by adjusting the soft masses of the third generation squark and the stop mixing $A_t$ so as to increase the Higgs mass while maintaining the same mass for $\tilde{t}_1$.  These modifications would have little impact on the rest of the analysis since it relies mostly on production channels which involve the first and second squark generations. 

\begin{figure}[hp]
\includegraphics[width=8cm]{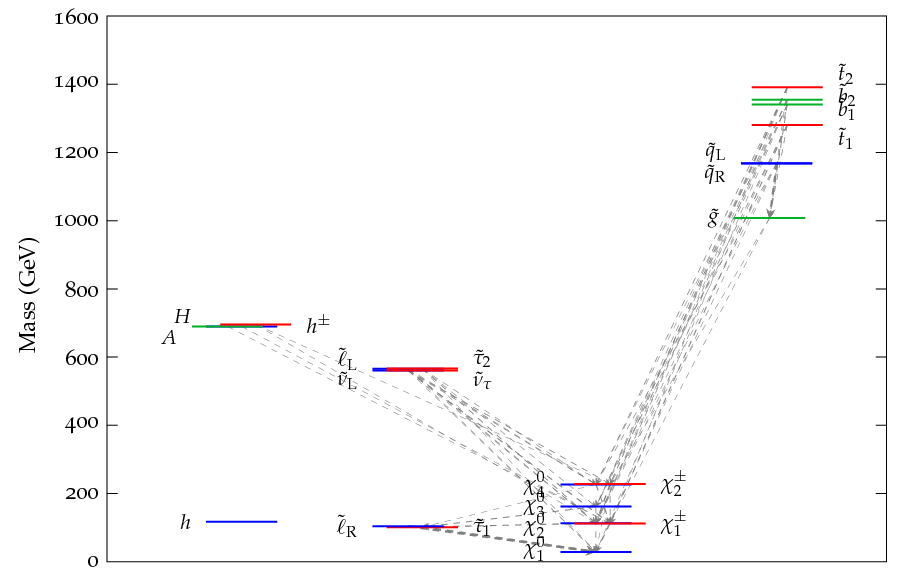}
\includegraphics[width=8cm]{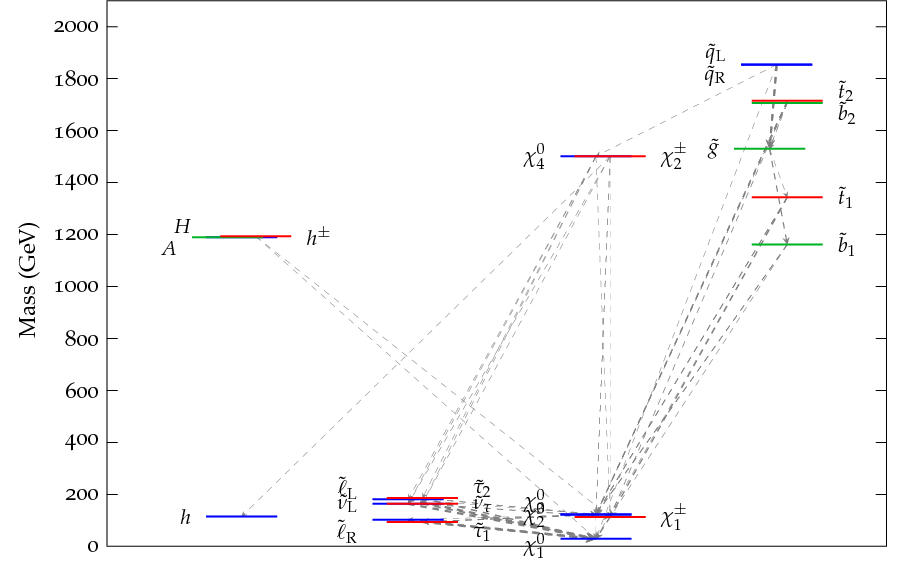}
\caption{Decay modes for BP-1 (left panel) and BP-2 (right panel)}
\label{fig:spectrum}
\end{figure}

BP-1 features masses for the first and second generation squarks and for the gluino which are just above the exclusion limit from ATLAS and CMS ~\cite{Aad:2011ib} and has gluinos lighter than all the squarks. The right-handed squarks have a large branching ratio into gluinos (87\%) while the left-handed squarks have significant branching ratios into $\tilde \chi_i q$ and $\tilde\chi^+ q'$.   The gluinos can therefore decay predominantly into charginos and neutralinos, according to the decay chain $\tilde{g}\rightarrow q\bar{q}\tilde\chi_i^0$ and
$\tilde{g}\rightarrow q\bar{q'}\tilde\chi_i^+$.  BP1 thus favour di and tri-tau production since neutralinos have  a large decay branching ratio into tau pairs and charginos into single taus (see Table~\ref{tab:1}). 
The full decay pattern for BP-1 is illustrated in Fig.~\ref{fig:spectrum}.

BP-2 features a much heavier coloured spectrum, which will naturally lead to suppressed cross sections for squarks and gluino production. 
In BP-2, the first and second generation squarks are heavier than the gluino and therefore decay directly into a quark and a gluino. The latter decays  as $\tilde{g}\rightarrow \tilde{b}_1 b$ with a branching ratio close to 90$\%$.  $\tilde{b}_1$  decays further  into quarks and neutralino/chargino, namely $\tilde{b}_1 \r t\tilde\chi_1^-, b\tilde\chi_2, b\tilde\chi_3$; each of these decay having a large branching ratio into taus.

\section{Collider Simulation}

We can now study the collider signatures of these two benchmark points.
As discussed earlier, the fact that the lightest chargino and the second lightest neutralino are light and dominantly Higgsino-like enhances the branching ratios into $\tilde\chi^{\pm}_1\r\stau_1\nu_{\tau}$ (50 $\%$-66 $\%$) and $\neu_2\r\stau_1\tau$ ($\sim 80\%$) respectively.
Therefore, staus (and taus) will be produced in cascade decays of squarks and gluinos via the decay of charginos and neutralinos, in addition to electroweak processes such as direct production of charginos/neutralinos and pairs of staus.  The signal has contribution mostly from the squark and gluino cascades, not only because of their high production rates but also because one gets harder taus which have better chances of surviving the cuts. The same-sign di-tau events result from the Majorana character of the gluino which decays into the $\tilde\chi^+_1$ and the  $\tilde\chi^-_1$ with equal rates, and charginos of the same sign in the two opposite hemispheres can yield a same-sign tau pair. Tri-tau events have their origin in the production of a $\tilde\chi^0_2$ and a $\tilde\chi^{\pm}_1$ in the two chains of cascade and
also in $\tilde{\chi}^0_2-\tilde{\chi}^0_2$ production in SUSY cascades, where both the neutralino decay into stau-tau pairs. The staus further decay into $\tilde{\chi}^0_1-\tau$ pairs, giving rise to four taus in
the final state. However due to limited tau identification efficiency, one of them goes untagged.

\subsection{Event signatures}

We will focus on the following processes:

\begin{itemize}
\item A pair of same-sign ditau-jets (SSD$\tau_j$), together with at least three hard central jets and large $\sla{E_T}$
(SSD$\tau_j+3-jets+\sla{E_T}$)

\item Three tau-induced jets in association with two/three hard central jets and large $\sla{E_T}$
($3\tau_j+3-jets+\sla{E_T}$).
\end{itemize}

\noindent{
Here, $\tau_j$ represents a jet out of a one-prong hadronic decay of tau, and the missing transverse energy is denoted by $\sla{E_{T}}$. 
The collider simulation has been done with a centre of mass energy $E_{cm}$=14 TeV,  using the event generator {\bf \begin{footnotesize} PYTHIA 6.4.16\end{footnotesize}} \cite{Sjostrand:2006za}. 
Once the Monte Carlo calculation yields a 
  cross-section, the number of events for any luminosity can be trivially
  obtained. Jets have been defined using the 
simple cone algorithm of PYCELL in PYTHIA with $\Delta R = 0.4$ (defined as $\Delta R=\sqrt{\Delta\eta^2+\Delta\phi^2}$, where $\eta$ is the 
pseudo-rapidity and $\phi$ is the azimuthal angle). We have used
the parton distribution function CTEQ6L1 \cite{Lai:1999wy} with the factorisation ($\mu_R$) and renormalisation ($\mu_F$)
scale set at $\mu_R =\mu_F =$ {\it average mass of the final state particles}. Initial and final state radiations have been taken into account. The finite detector resolutions have also been incorporated according to the prescription 
given in ~\cite{Biswas:2009zp}. Following are the numerical values of various parameters, used in our calculation \cite{Amsler:2008zzb}:}
\\

$~~~~~~~~$ $M_Z=91.187$ GeV,   $M_W=80.398$ GeV,   $M_t=171.4$ GeV  \\
$~~~~~~~~~~~~~$ $\a^{-1}_{em}(M_Z)=127.9$, $~~$  $\a_{s}(M_Z)=0.118$\\

We have considered one-prong hadronic decay of tau's, which comprise 80\% of its hadronic decay width and about
50\% of its total decay width. In the massless limit ($E_{\tau}\gg m_{\tau}$) where the tau is boosted in the 
laboratory frame, tau decay products are nearly collinear with the parent tau. In this limit, hadronic tau decays 
produce narrow jets of low multiplicity, to be identified as tau-jets. The one-prong hadronic decay of the tau has 
been identified using the prescription of ~\cite{tauid}, assuming a true tau-jet identification efficiencies of 50\%
and a fake tau-jet rejection factor of 100 ~\cite{tauid,taueff} for both signal and backgrounds. We have also assumed 
that for a true tau-jet, the charge identification efficiency is 100\%, while to a non-tau jet we have randomly assigned 
positive and negative charge, each with 50\% weight \footnote{The charge identification efficiency in one-prong hadronic 
decay of tau is $\sim85\%$ \cite{tauid}. However, the probability that a background 
jet will show up as a one-prong tau-jet is also small (60\%-70\%) rather than our conservative assumption of 100\%.}. 
Note that we did not consider leptonic decays of the taus and this
has been ensured by vetoing isolated leptons in the final state. In Table \ref{tab:xsec} we present the initial hard scattering 
cross-sections for strong and electroweak-gaugino productions. Though the EW gaugino production cross-section dominates over 
the strong production, in the following one can see, after applying the background elimination cuts, it is the strong production 
which dominates the signal cross-section.

\begin{table}[htbp]
\begin{center}
\begin{tabular}{|l|c|c|c|c|}
\hline
\hline
Benchmark &\multicolumn{4}{c|}{Processes} \\
 ~~ Points   & $\tilde{g}\tilde{g}$ & $\tilde{q}\tilde{q}^{(*)}$ & $\tilde{q}\tilde{g}$  &   EW gaugino  \\
\hline
BP1   & $0.7191$  &  $0.3055$  & $0.5053$ &  $11.867$ \\
\hline
BP2   &  $8.949\times10^{-3}$ & $2.4036\times10^{-2}$   & $1.7256\times10^{-2}$ & $8.088824$ \\
\hline 
\hline
\end{tabular}
\caption{\small \it {The hard scattering cross-sections (in pb) of strong and electroweak productions
for our respective benchmark points.}}
\label{tab:xsec}
\end{center}
\end{table}


\subsection{Standard model backgrounds }

Both of our final states suffer from standard model (SM) contaminations. The following SM processes are the most important ones.

\begin{itemize}

 \item {\bf $t\bar{t}$}: This is potentially the most serious among all the SM backgrounds. This contributes to the
$SSD\tau$ final state in various ways. One can have two same sign taus one from a W produced in top (anti-top) 
decay and the other from a b-quark produced in anti-top (top) decay. One can also have a tau from the decay of any 
one of the W or b and a jet can be faked as a tau to give a same sign di-tau final state. It can also happen that any 
two of the non-tau jets produced in $t\bar{t}$ events can be faked as tau-jets. Similarly, a tri-tau final state can also 
be mimicked if two taus come from the decays of the two W's and one from b-decay or one from W-decay and the other two
from the decays of the two b-quarks. In addition, one can have two true tau-jets from either the two W's or two b's
or one each from W and b-decay while the third one can be a non-tau jet faked as a tau. A very small fraction of events
contain one true-tau jet from either one of the two W's (or b's) and two non-tau jets faked as tau-jets. Though the efficiency of 
a non-tau jet being identified as a narrow tau-like jet is small, the overwhelmingly large number of $t\bar{t}$ events produced at 
the LHC makes this subprocess  a dominant source of background.

\item {\bf $t\bar{t}$}+jets: This subprocess contributes to our final state in a similar way as $t\bar{t}$ does. The only difference
is that with additional jets in the final state the jet faking probability increases in this case. Apart from this, the kinematics
of these events are slightly  different from that of $t\bar{t}$ events, which show up in the distributions of various kinematic variables.

\item {\bf $ZZ$}: This subprocess is also a potential background to our desired final state due to the limited tau identification efficiency. The case when both the Z decays into $\tau^+\tau^-$-pairs 
($ZZ\r\tau^+\tau^-\tau^+\tau^-$) and two of the tau-jets  having same sign get detected while the other two go undetected
 contribute to $SSD\tau_j$ final state.
 Also, if one of the Z decays into a $\tau^+\tau^-$-pair and the other decays
hadronically, then the combination of one true tau-jet and one fake tau-jet can give rise to same sign di-tau jet final state. A tri-tau jet 
final state can be mimicked when one  fake tau-jet is identified in addition to the tau pairs from Z decays. One can also
have one true tau-jet out of a Z-decay and any two non-tau jets being identified as tau-jets which can then fake the tri-tau jet final state. 

\item {\bf $ZH$}: This background is similar to  the case of $ZZ$. Here, the decay of Higgs ($H\r \tau^+\tau^-$) can produce tau lepton in the final state. The 
non-tau jets can arise in $H\r b\bar{b}$.

\item {\bf $ZW$}: This subprocess can contaminate both di-tau jets and tri-tau jets final state. The same sign di-tau can arise when one tau from Z-decay
and the other one from W-decay get identified. Also one can have a true tau-jet either from Z or W decay and the other one can be a fake tau-jet.
Contribution to the tri-tau jet final state comes from events where both the Z and the W decay in the tau-channel. The combination of two (one) true tau jets and one (two)
fake jet(s) is also a possibility in this case.


\end{itemize}

We have simulated $10^7$ signal events using PYTHIA for both of our benchmark points. The $t\bar{t}$ and $t\bar{t}$+jets backgrounds have
been simulated using ALPGEN (version 2.14) \cite{Mangano:2002ea}. We have generated $10^8$ weighted events for $t\bar{t}$+0-jet (and $t\bar{t}$+1-jet)  
and left with 2303879 (and 638838) unweighted events, whereas for $t\bar{t}$+2-jets (and $t\bar{t}$+3-jets) we have generated $5\times10^8$ weighted events  
and ended up with 516567 (and 306035) unweighted events. The remaining $ZZ$, $ZH$ and $ZW$ processes have been simulated using PYTHIA generating $10^7$
events. In all cases, the initial and final state radiation and the parton hadronization are simulated by PYTHIA.

\subsection{Event selection criteria}

We have implemented the following basic cuts for each event to validate our desired final states. For the
SSD$\tau$ final state we have used the following event selection criteria:

\begin{itemize}
\item  $p_{T_{\tau_1}} >$  50 GeV and $p_{T_{\tau_2}} >$  40 GeV, $|\eta| <$  2.5 
for the two tau-jets in the final state.
\item $p_T >$  100, 100, 50 GeV, $|\eta| <$  2.5 for the three associated jets, in decreasing 
order of hardness. 
\end{itemize}

The pre-selection cuts imposed for the 3$\tau$ final state are:

\begin{itemize}
\item  $p_{T_{\tau_1}} >$  40 GeV, $p_{T_{\tau_2}} >$  30 GeV and $p_{T_{\tau_3}} >$ 30 GeV, $|\eta| <$  2.5 
for the three tau-jets in the final state.
\item $p_T >$  100, 75, 50 GeV, $|\eta| <$  2.5 for the three associated jets, in decreasing 
order of hardness. 
\end{itemize}

\begin{figure}
\includegraphics[width=0.48\textwidth]{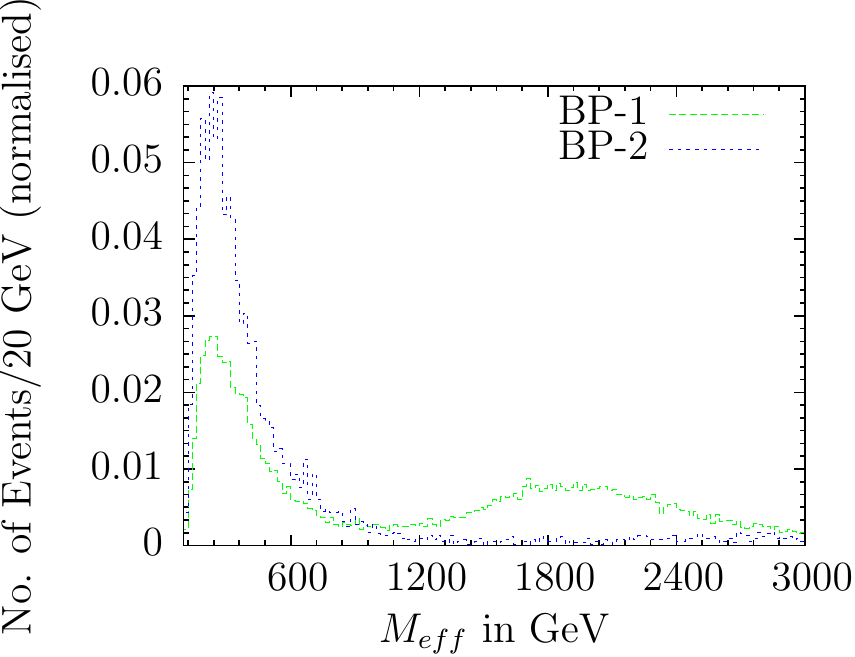}
\includegraphics[width=0.48\textwidth]{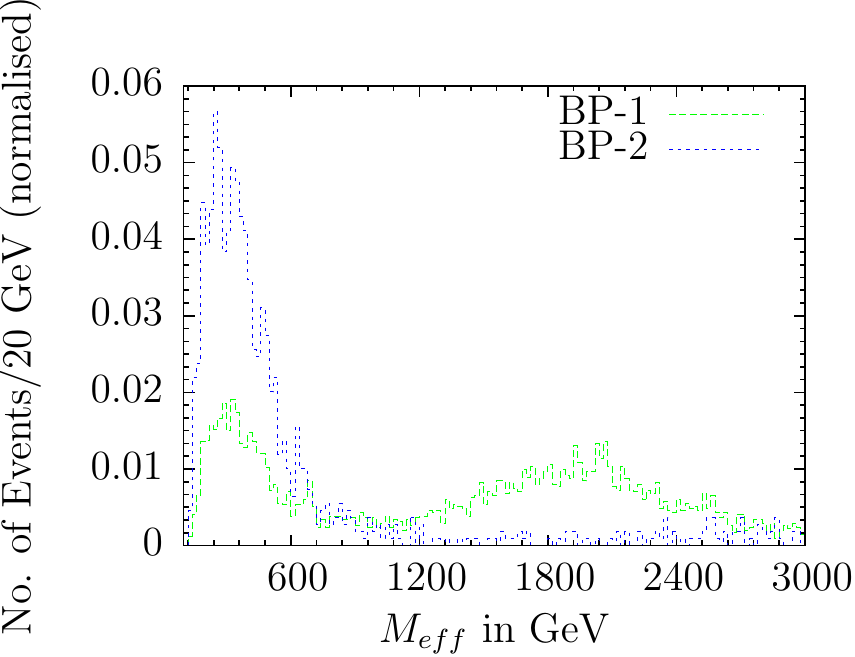}
\includegraphics[width=0.48\textwidth]{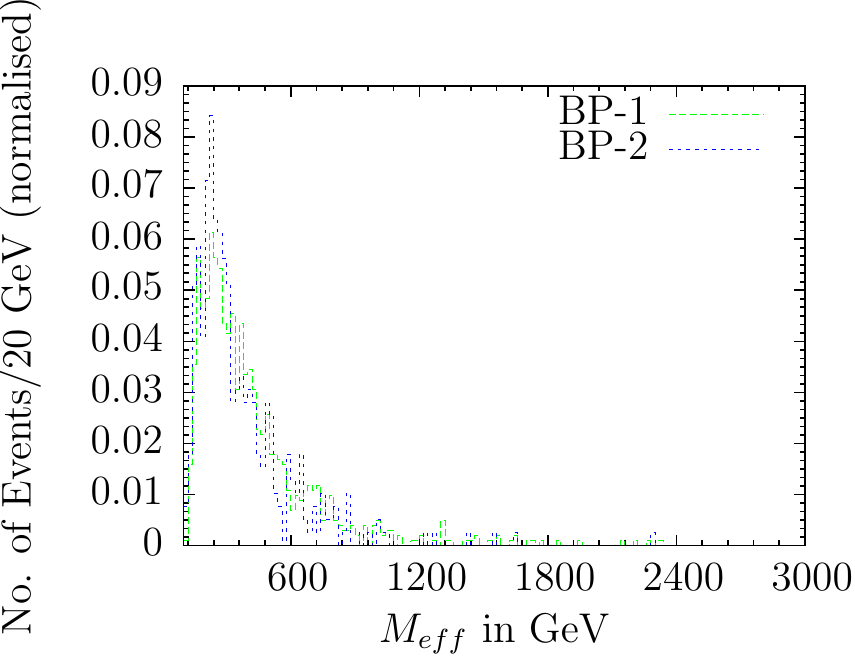}
\includegraphics[width=0.48\textwidth]{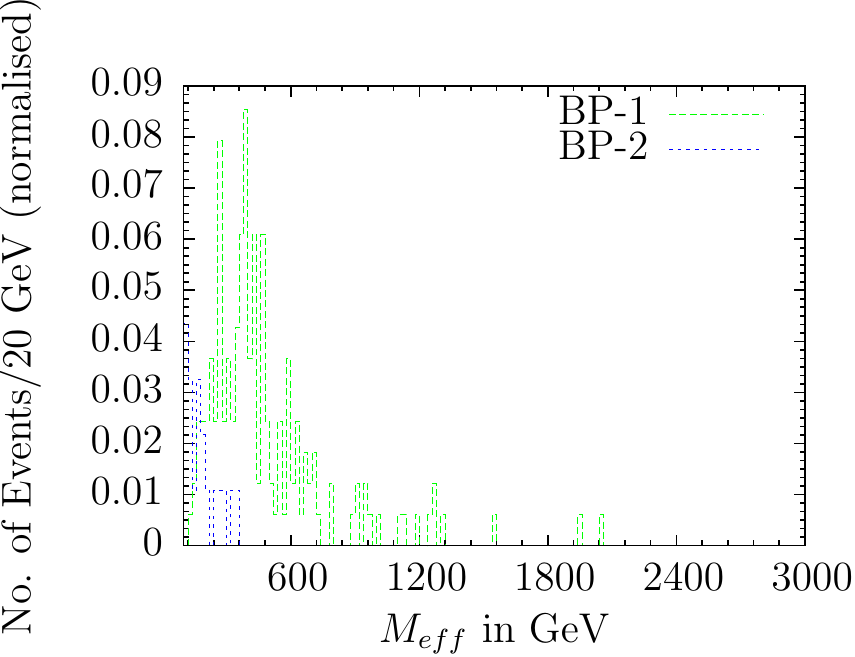}
\caption{Effective mass reconstruction 
for di (left) and tri-tau (right) channels. 
The two figures above illustrate the normalised distribution for tau production including all SUSY process for our two benchmark points while the two figures below show 
the tau production in which we include only the electroweak gauginos and $\tilde{\tau}_1$ $\tilde{\tau}_1$* production. }
\label{meff}
\end{figure}

\begin{figure}
\includegraphics[width=0.48\textwidth]{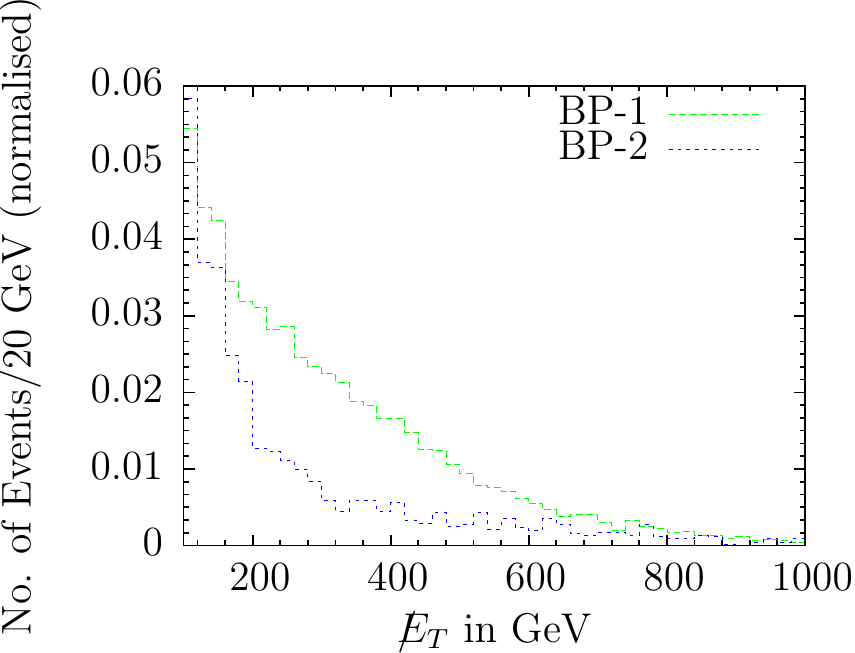}
\includegraphics[width=0.48\textwidth]{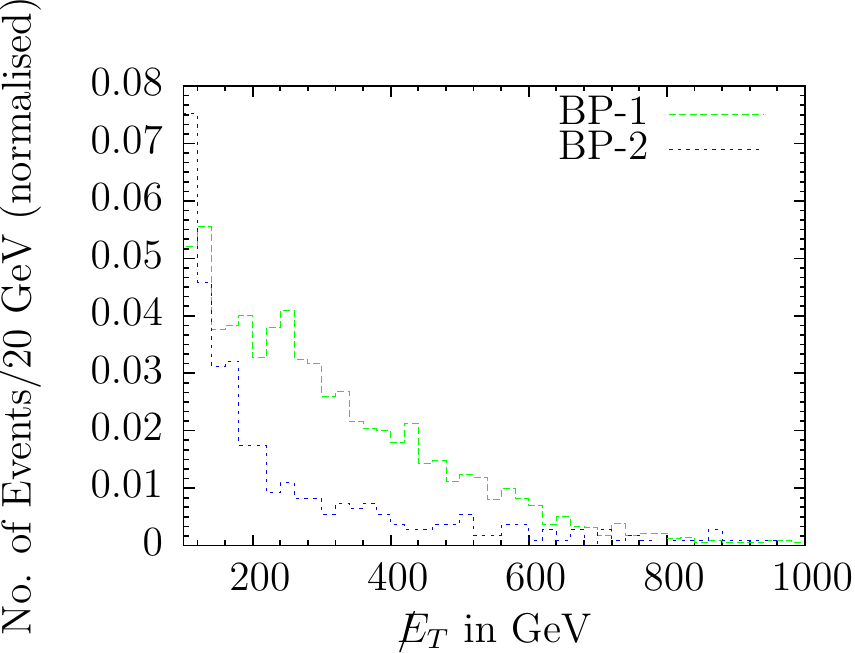}
\includegraphics[width=0.48\textwidth]{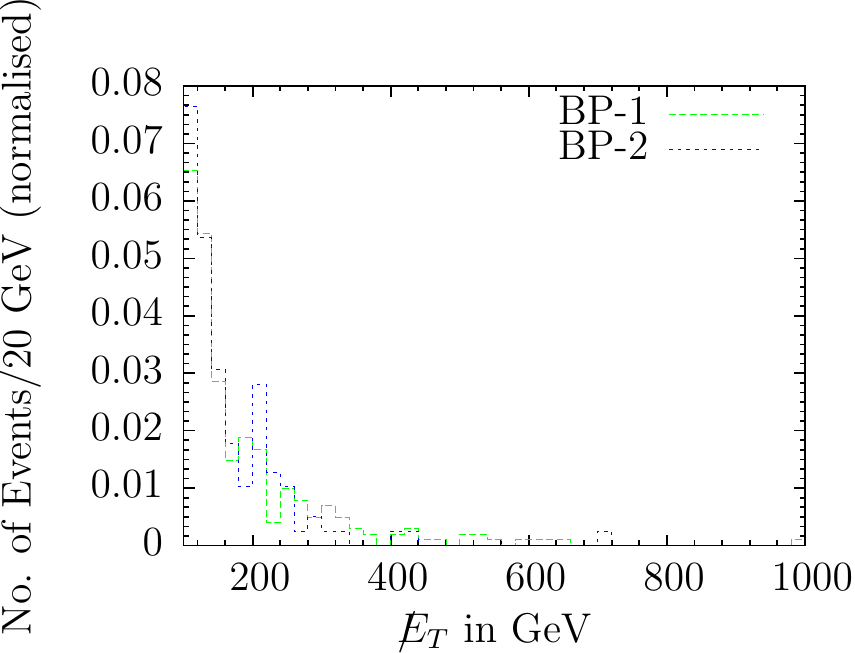}
\includegraphics[width=0.48\textwidth]{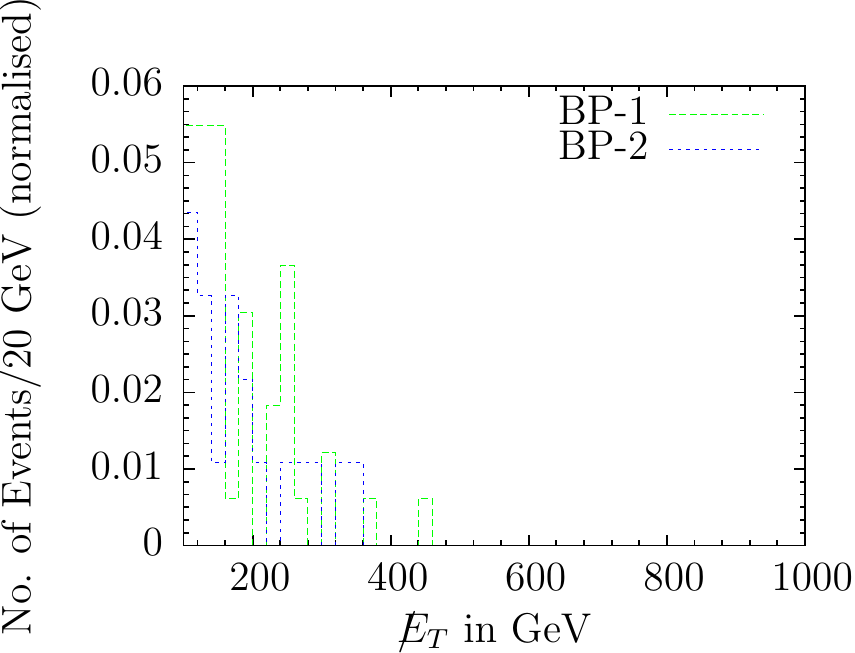}
\caption{Normalised  missing energy distribution for di (left) and tri-tau (right) channels. 
The two figures above illustrate the tau production including all SUSY process for our two benchmark points while the two figures below show 
the tau production in which we include only the electroweak gauginos and $\tilde\tau_1\tilde\tau_1^*$ production. }
\label{met}
\end{figure}

\begin{figure}
\includegraphics[width=0.48\textwidth]{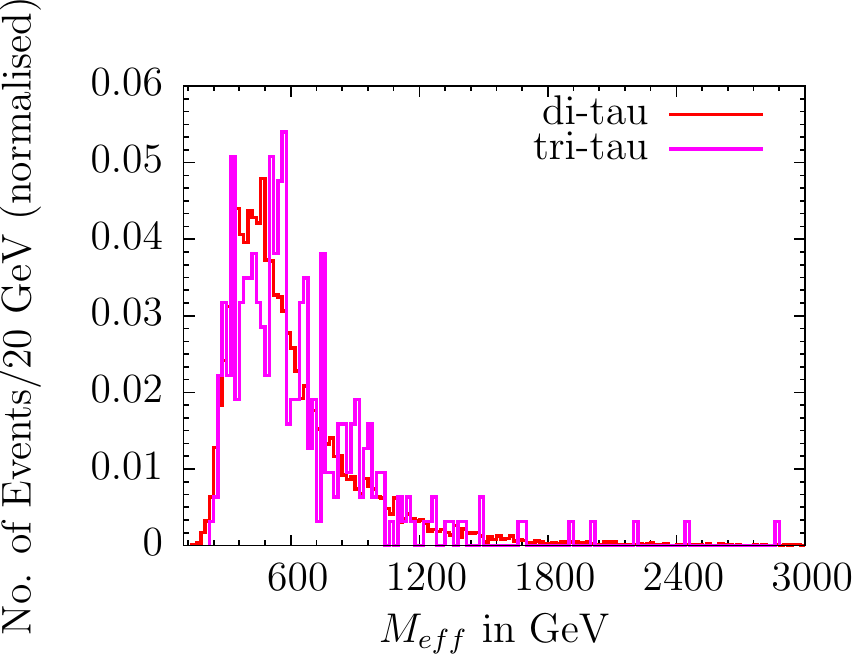}
\includegraphics[width=0.48\textwidth]{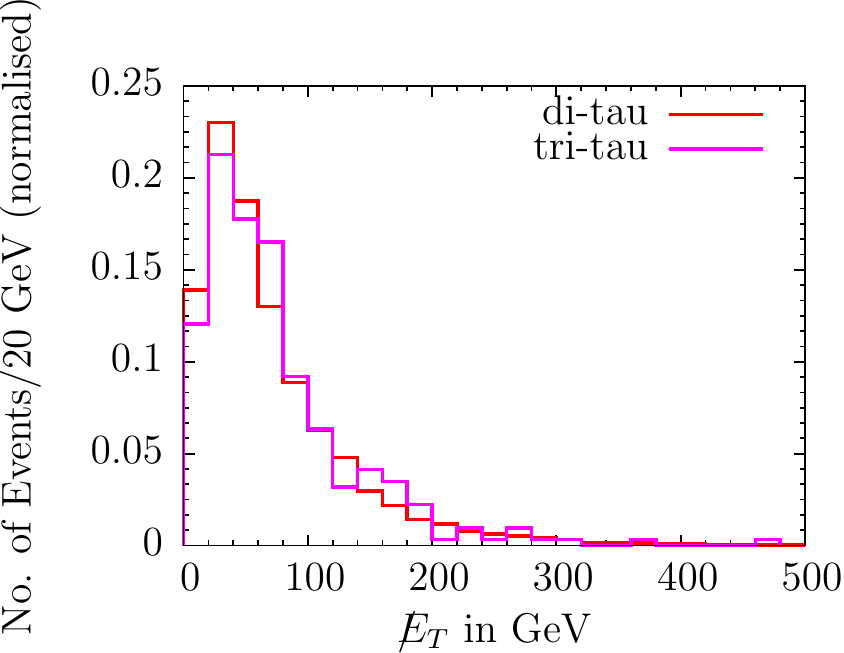}
\caption{Normalised effective mass (left) and missing energy (right) distribution for the di and tri-tau channels. 
The two figures above illustrate the standard model background contributions to our desired final states. }
\label{backg}
\end{figure}

To reduce the standard model background we have investigated  additional cuts on the following 
kinematic variables, as can be motivated from figure \ref{meff}, \ref{met} and \ref{backg}. These 
cuts considerably suppress the SM backgrounds with less severe affect on the signal,

\begin{itemize}
\item $\Sigma|\vec{p}_T|>$ 1000 GeV where $\Sigma |\vec{p}_T|$ is the scalar sum of the transverse
momentum of all visible particle in the final state.
\item $\sla{E}_T >$max(150,~0.1$\cdot\Sigma|\vec{p}_T|$)
\end{itemize}

or 

\begin{itemize}
\item $M_{eff}>$ 1000 GeV where, $M_{eff} = \Sigma|\vec{p}_T|+\sla{E_T}$.
\item $\sla{E}_T >$max(150,~0.1$\cdot M_{eff}$)
\end{itemize}

Those cuts were selected to give the best significance, a comparison of the impact for different values and combination of cuts 
will be presented in the next section. Note that the Z-invariant mass cut will reduce the background from Z+jets considerably, 
and the application of the hard cut on scalar sum of $p_T$ and/or effective mass variable reduces both the QCD and Z+jets backgrounds, 
thus the dominant backgrounds are mainly $t\bar{t}$ and $t\bar{t}$+jets.

\section{ Results and discussions}

In this section we present the numerical results of our analysis. Table \ref{tab:events} shows the expected
signal and background events at the 14 TeV LHC run and with  an integrated luminosity of 
10 fb$^{-1}$. The following estimator for the signal significance is used ~\cite{tauid}

\be
Sig. = \sqrt{2((S+B){\rm ln} (1+ S/B)-S)}
\ee
\noindent{where, S and B are the expected number  of signal and  background events, respectively.}

First note that $\tilde{q} \tilde{q}*, \tilde{q} \tilde{g}, \tilde{g} \tilde{g}$ are the dominant processes  for the BP-1 di-tau production after 
application of basic cuts while they are not for BP2 because the gluinos and squarks are too heavy. As a result electroweak gaugino decay lead to a 
substantial contribution (~50 $\%$) in the di and tri-tau BP-2 production channels. However, after applying various cuts to suppress the background, 
only the contributions from $\tilde{q} \tilde{q}*, \tilde{q} \tilde{g}$ (and $\tilde{g} \tilde{g}$ for BP1) will survive so that  the significance will 
be large only for BP-1.

\begin{table}[htbp]
\begin{tabular}{|l|c|c|c|c|c||c|c|c|c|c|}
\hline
\hline
 Cuts &\multicolumn{5}{c||}{\bf SSD$\tau_j$}&\multicolumn{5}{c|}{\bf Tri$\tau_j$}\\
  &  B & BP1 & Sig. & BP2 & Sig. & B & BP1 & Sig. & BP2 & Sig. \\
\hline
basic cuts &  2368 & 355 & 7.12 & 39 &   0.799     & 138  & 82 & 6.41 &  14 & 1.17 \\
\hline
$\sla{E_T}>150$ GeV  &  376 &  259 & 12.15  & 22 &  1.12     & 19   & 60 & 10.25  &  8 & 1.72 \\
\hline
$\Sigma{|p_T|}>1000$ GeV & 482 & 294 & 12.29 & 19 & 0.86      &  18  & 69 & 11.67 &  7 &  1.56 \\
\hline
$\Sigma{|p_T|}>1100$ GeV & 319 & 280 & 13.96 & 19 & 1.05       & 12  & 67 & 12.79 & 7  & 1.86 \\
\hline
$M_{eff}>1100$ GeV   & 326 &  296 & 14.55 & 19 &  1.04            &  14  & 69 & 12.55  & 7 & 1.74 \\
\hline
$M_{eff}>1200$ GeV   & 257 & 287  & 15.5  & 19 &  1.17              &  10  & 68  & 13.58  & 7 & 2.01 \\
\hline
$\Sigma{|p_T|}>1000$ GeV+$\sla{E_T}>$ & 106 & 208 & 16.31 & 15 & 1.42      &  8  & 52 & 11.74 &  7 & 2.20 \\
$max(150,~0.1\Sigma{|p_T|})$ GeV  &&  &  &  &               &&  &  &  & \\
\hline
$M_{eff}>1000$ GeV+$\sla{E_T}>$ & 157 & 246 & 16.36 & 19 & 1.49              &  10  & 58 & 12.03 &  8 & 2.27 \\
$max(150,~0.1M_{eff})$ GeV  &&  &  &  &                  &&  &  &  & \\
\hline 
\hline
\end{tabular}
\caption{\small \it {Number of signal and background events for the $2\tau_j$+3-jets+$E_{T}\miss$ 
and $3\tau_j$+3-jets+$E_{T}\miss$ final states, considering all SUSY processes, with $E_{cm}$=14 TeV 
at an integrated luminosity of 10 $fb^{-1}$ assuming tau identification efficiency of 50\% and a jet 
rejection factor of 100. The series of cuts are applied independently.}}
\label{tab:events}
\end{table}

As we can see from the upper panel in Table~\ref{tab:events}, both the di and tri-tau production channels are larger
for BP-1 at large effective mass values ($M_{eff} > 1200$ GeV)  than for BP-2. For these values the signal is dominated by squark 
and gluino initiated processes and is therefore larger for lighter  squarks and gluinos.
For  BP-2, there is a larger signal  at low effective mass values ($M_{eff} < 600$ GeV) than for BP-1. In this region the  
signal is completely dominated  by  electroweak, i.e. charginos,neutralinos and staus, contributions  
(see lower panel). Thus a kinematic cut at large $M_{eff}$ (i.e. above 1200 GeV) would not   only  get rid of the background but 
also most of the signal from electroweak processes. This means that such a cut would unfortunately kill  most the signal for BP-2.  
On the other hand, for BP-1,  the signal significance is large enough to discover supersymmetry in di and tri-tau production.
 
The missing $E_T$ distribution is more peaked towards values for BP-2 than for BP-1 because BP-2 is dominated by electroweak processes,
as can be seen from a comparison of the top and bottom panel of Fig.~\ref{met}. A hard cut on MET will therefore enhance  mostly the significance for BP-1. The distributions for the scalar sum of transverse momentum have similar shapes as the $M_{eff}$ distributions. 
The impact of the various cuts are listed in Table \ref{tab:events} where one sees  that harder cuts lead to large significance for BP-1 in both the di-tau and tri-tau channels.

Because of the low number of signal events for BP-2, a higher luminosity is required to reach the discovery level in both channels. We estimate that the significance would increase by a factor 3 for a luminosity of ${\cal L}=100 {\rm fb}^{-1}$.

\section{Summary and conclusions}

We have investigated the LHC signatures of an MSSM scenario which
addresses the dark matter issue with the help of neutralinos that are rather
on the lightest side, namely, below 30 GeV. We have taken a cue from an earlier
study which indicated that such a situation,  for its consistency, will also require
light  non-strongly interacting superpartners, and particularly a stau as the
next-to-lightest SUSY particle. The exchange of light stau in this case is required
for achieving the desired annihilation rate for the dark matter candidate.
Accordingly, we have focused on the signals
where such a light stau has the central role, and in this spirit we have
explored same-sign ditau and tritau events.
It should, however,  be noted that our analysis can also be applied
 to other SUSY scenarios where relatively light  staus are abundantly
produced in SUSY cascades.

We have chosen two benchmark points representative of the scenario
under investigation. These benchmark points comply with all astrophysical
as well as accelerator constraints, and they predict large branching ratios
for the chargino and (second lightest) neutralino decaying into staus.
As the stau-rich  final states are susceptible to backgrounds of a challenging
nature, a study of the relevant backgrounds has also been performed.
It is found that one has to wait for the 14 TeV run, and integrated
luminosities ranging from 10 to 100 fb$^{-1}$ in order to see the signals
with reasonable statistical significance.   

Our results show that, since both the di-and tri-tau production processes
peak at large effective mass for benchmark point 1 (where strong production
controls the rates), a sizable effective mass cut is rather useful in removing backgrounds.
This cut is not so efficient for benchmark point 2, since the events there arise
largely  from electroweak production. That is why higher luminosities are
required for the effective exploration of this region of the parameter space.
This conclusion is one more reminder of the general difficulty in exploring
MSSM scenarios where the squarks and gluinos are heavy but the non-strongly
interacting new particles are quite light. The difficulty is compounded when one is
looking for smoking gun signals of a specific situation within the above class,
namely, where a light stau has a special role in ensuring the requisite
relic density. Our analysis shows that such a scenario can be probed at the LHC
in spite of all difficulties, provided the LHC runs through sufficient
integrated luminosity at 14 TeV. On the other hand, any non-observation of the di-tau and the 
tri-tau signal rates over the SM backgrounds at the LHC would highly constrain this 
type of scenarios.

\section*{Acknowledgements} 
BM acknowledges the hospitality of Fawzi Boudjema and LAPTH where this project
  was initiated.
  The work of BM has been partially supported by funding available for the Regional Centre for
  Accelerator-based Particle Physics from the Department of Atomic Energy, Government of India. 


\end{document}